\documentclass{aa}  
\usepackage{float}
\usepackage{graphicx}
\usepackage{txfonts}
\usepackage{hyperref}
\begin{document}

   \title{Transit of asteroids across the 7/3 Kirkwood gap under the Yarkovsky effect}

   \author{Yang-Bo Xu\inst{1,2}
          \and
          Li-Yong Zhou\inst{1}\fnmsep\inst{2}\fnmsep\inst{3}
          \and
          Wing-Huen Ip\inst{3,4}
          }
   \authorrunning{Xu, Zhou \& Ip}
   \offprints{Zhou L.-Y., \email zhouly@nju.edu.cn}
   \institute{School of Astronomy and Space Science, Nanjing University, 163 Xianlin Avenue, Nanjing 210046, China
         \and
            Key Laboratory of Modern Astronomy and Astrophysics in Ministry of Education, Nanjing University, China
          \and
       Institute of Space Astronomy and Extraterrestrial Exploration (NJU \& CAST),  Nanjing University, China
        \and
     Institute of Astronomy, National Central University, Jhongli, Taoyuan City 32001, Taiwan}

   \date{}

  \abstract
  {Yarkovsky effect plays an important role in asteroids drifting in the inner Solar System.  In the main belt, many asteroids are continuously pushed by Yarkovsky effect into regions of different mean motion resonances (MMRs) and then ejected after a period of time due to the instability of MMRs. They are considered as the principal source of near-Earth objects. We investigate in this paper by numerical simulations the effects of the 7/3 MMR with Jupiter (J7/3 MMR) on the transportation of asteroids from Koronis family and Eos family that reside respectively on the inner and outer side of the resonance. The J7/3 MMR acts like a selective barrier to migrating asteroids. The fraction of asteroids that make successful crossing through the resonance and the escaping rate from the resonance are found to depend on the Yarkovsky drifting rate, the initial inclination and the migrating direction. The excitation of eccentricity and inclination due to the combined influence from both the resonance and Yarkovsky effect is discussed. Only the eccentricity can be pumped up considerably, and it is attributed mainly to the resonance. In the observational data, family members are also found  in the resonance and on the opposite side of the resonance with respect to the corresponding family centre. The existence of these family members is explained using our results of numerical simulations. Finally, the replenishment of asteroids in the J7/3 MMR and the  transportation of asteroids by it are discussed. }


   \maketitle
%

\section{Introduction}
Yarkovsky effect produces a non-gravitational force that arises from the absorption and anisotropic re-radiation of thermal energy. An asteroid receives the radiation from the Sun and re-radiates the energy out after a while. Because of the rotation and revolution of the asteroid, the net radiation pressure leads to a tiny recoil force, and the accumulated effect drives the semimajor axis of the asteroid to increase or decrease continuously. With the  complete linear model given by \citet{1999A&A...344..362V}, the Yarkovsky effect has been introduced to solve problems in many aspects, such as the dispersion of asteroid family members and the origin of near-Earth objects (NEOs) \citep[see e.g.][for a review]{2006AREPS..34..157B}. 

The resonant zone in the main belt in between orbits of Mars and Jupiter is considered to be the major source of NEOs. Asteroids were once thought to be ejected into resonances by collisions with each other. However, this mechanism shall deliver to the Earth a large fraction of meteoroids with cosmic-ray exposure (CRE) ages of a few million years \citep{1993Icar..101..174F,1997Sci...277..197G}, which is inconsistent with the CRE ages of most stony and iron meteorites \citep{1998M&PS...33..999M}. This disagreement remained unsolved until Yarkovsky effect was brought up \citep{1998Icar..132..378F}. Nowadays, it is widely accepted that the main belt asteroids are pushed into resonances mainly by Yarkovsky effect, and later their eccentricities may be excited by the resonances until close encounters with planets.
 
The 7/3 mean motion resonance (MMR) with Jupiter (hereafter J7/3 MMR) is one of the important resonances in the main belt despite its high order. Located at $a_{7/3}\approx 2.957\,$AU and with a width of about 0.01\,AU \citep{2006Icar..182...92V}, the J7/3 MMR is thought to be unstable and responsible for the Kirkwood gap around its position. \citet{1989A&A...213..436Y,1991Icar...92...94Y} found large eccentricity variations and chaotic behaviours of some asteroids in the J7/3 MMR, but the eccentricity did not necessarily vary greatly in the resonance. Later, the secular resonances $\nu_5$ and $\nu_6$, where the perihelion of an asteroid precesses at the same pace as Jupiter (or Saturn) does, were found in the region \citep{1995Icar..114...33M}, and the overlapping of $\nu_5$, $\nu_6$ and J7/3 MMR introduces chaos that dominates the motion in this region.  The eccentricity of asteroids in this chaotic region can be pumped up to 0.8 at most. The structure of the J7/3 MMR was also investigated by \citet{2003Icar..166..131T}, and the instability was confirmed again.

Affected by the Yarkovsky effect, an asteroid is able to drift freely in semi-major axis in the non-resonant region. However, the migration will be modified by resonances, especially the MMRs. Some asteroids may cross the resonance while some others may be trapped in. The J7/3 MMR is such a resonance, serving as a selective barrier for drifting asteroids. 

Two major asteroid families reside around the J7/3 MMR, namely Koronis family on the left hand side (with smaller semi-major axis) and Eos family on the right hand side. 
\citet{2006Icar..182...92V} have made a few simulations of Eos asteroids transiting the J7/3 MMR and given some qualitative conclusions, including that Eos family cannot extend to the other side of the J7/3 MMR. 
However, the classification by \citet{2014Icar..239...46M} showed that there are a few Eos members on the left hand side of the resonance in addition to those in the resonance (see Section 4).
\citet{2003Icar..166..131T} found that all asteroids in the J7/3 MMR have short dynamical lifetimes, so that there must be a continuous supplement of asteroids, for which they found that the diurnal Yarkovsky effect could provide a flux of members from both Koronis and Eos families enough for maintaining the population in the J7/3 MMR.  

As more and more asteroids have been identified as members of both families, it is desirable to revisit the transiting of asteroids across the J7/3 MMR in greater detail, as a coherent process, and in a full dynamical model. We report here our investigations on the transit of asteroids across the J7/3 MMR under the influence of Yarkovsky effect. The rest of this paper is organized as follows. In Sect.\,\ref{sect:simulation}, we introduce the dynamical model and initial setting of the numerical simulations. In Sect.\,\ref{sect:result}, the results of simulations will be presented. In Sect.\,\ref{sect:app}, we compare our results with observational data, and the supplement of asteroids into the J7/3 MMR is discussed. And finally, conclusions are summarised in Sect.\,\ref{sect:conclusion}.


\section{Dynamical model with Yarkovsky force} \label{sect:simulation}
To investigate the behaviour of asteroids transiting the J7/3 MMR, we numerically simulate their orbital evolutions. The dynamical model adopted in our simulations consists of the Sun, all planets but Mercury and massless fictitious asteroids (test particles). It is worthy to note that four terrestrial planets were generally ignored in the study of dynamics of main belt asteroids \citep[e.g.][]{2003Icar..166..131T}. We exclude only the innermost planet Mercury in our model because its small mass and short orbital period make its perturbation ignorable, and also to include Mercury requires much more computations. We adopt the zeroth correction of Mercury by adding its mass to the Sun. We adopt the Earth-Moon barycentre instead of the separate Earth and Moon as usual \citep[e.g.][]{2012A&A...541A.127D, 2019A&A...622A..97Z}. The initial orbits of the planets at epoch  JD\,2457400.5 are derived from the JPL's HORIZONS system\footnote{\url{ssd.jpl.nasa.gov/horizons.cgi}}\citep{1996DPS....28.2504G}.

To simulate the motion of asteroid members of both Koronis and Eos families near the J7/3 MMR, we generate two sets of test particles, one on the inner side of the resonance ($a_p<a_{7/3}$), and the other on the outer side. The inner set mimicking members of Koronis family is labelled $Kor_{in}$ after Koronis hereinafter, and the outer set (with $a_p>a_{7/3}$) $Eos_{out}$ after Eos family by the same token. Each set includes 1000 test particles, and we will check how they transit through the J7/3 MMR under the influence of Yarkovsky effect.

The initial positions of these test particles are chosen carefully. A test particle too far away from the J7/3 MMR will waste plenty of integration time on its way approaching to the resonance, while too close to the J7/3 MMR will miss the process of drifting into the resonance from the non-resonant region. After some attempts, we set the initial semimajor axis $a_0=2.948\,$AU for $Kor_{in}$ and $a_0=2.962\,$AU for $Eos_{out}$. 

The proper eccentricities $e_p$ of Koronis family and Eos family are both approximately in the range from 0.04 to 0.10. They have different proper inclinations $i_p$, averagely $2^\circ$ and $10^\circ$ for Koronis and Eos family, respectively \citep{2003Milani}.
Following the same strategy applied in \citet{2003Icar..166..131T}, we set the initial osculating eccentricities $e_0$ randomly vary from 0.08 to 0.14 and initial osculating inclinations $i_0=3^\circ$ for $Kor_{in}$ and $i_0=11^\circ$ for $Eos_{out}$. These numbers are used just to make the test particles have the same proper elements as the asteroids of these two families. The rest angular elements, including longitude of the ascending node $\Omega$, argument of perihelion $\omega$ and mean anomaly $M$, having not long-term influence on the orbital evolution, are set arbitrarily $\Omega_0=100^\circ$, $\omega_0=274^\circ$ and $M_0=285^\circ$. 

In addition, to investigate the resonance transit in the opposite direction, we make other two sets of test particles labelled $Kor_{out}$ and $Eos_{in}$, which are the same as $Kor_{in}$ and $Eos_{out}$ except for the initial semi-major axes. $Kor_{out}$ and $Eos_{in}$ are placed at $a_0=2.962\,$AU and $a_0=2.948\,$AU, respectively.

The recoil force of Yarkovsky effect is dependent on a series of physical parameters, most of which have not been measured accurately yet. As a compromise, Yarkovsky effect is generally degenerated into an equivalent drift rate of semi-major axis $\dot{a}_Y$.  For a ``typical'' asteroid at $a=2.9\,$AU with radius $R=1\,$km and rotation period $P=8\,$hrs, assuming typical parameters as follows \citep{2006Icar..182...92V}: albedo $p=0.13$, thermal conductivity $K=0.005\,$W/m/K, specific heat capacity $C=680\,$J/kg/K, surface density $\rho_s=1.5\,$g/cm$^3$, bulk density $\rho=2.5\,$g/cm$^3$ and spinning obliquity $\gamma=0^\circ$, its Yarkovsky drift rate can be calculated, $\dot{a}_Y=0.128\,$AU/Gyr \citep{2019Xu,2019A&A...622A..97Z}. The currently known smallest asteroids of Koronis family and Eos family have radii of about 0.5\,km, i.e. half of the above mentioned example asteroid, whose Yarkovsky drifting thus will be twice quicker $|\dot{a}_Y|=0.256\,$AU/Gyr. Considering the uncertainties of parameters, the maximal drift rate may be larger than this value, while the minimum can be as small as 0 (e.g. if $\gamma=90^\circ$). To make our simulations be able to represent all possible situations, we set 8 different drifting rates $|\dot{a}_Y|=0.02, 0.05, 0.08, 0.1, 0.2, 0.3, 0.4, 0.5\,$AU/Gyr. In general, most of asteroids affected by Yarkovsky effect will be in this domain of migration speed.

Our aim is to investigate the orbital evolution of test particles when they are crossing through the J7/3 MMR region under the influence of Yarkovsky effect, thus the timespan of our simulation $T$ should be long enough to allow the migration of asteroid into and out of the resonance to complete. However, asteroids may ``wander'' in the resonance for a while rather than cross it immediately without any hesitation. The time delay in the resonance can be estimated using the functional dependence of the trapping time in the resonance on $|\dot{a}_Y|$ and the strength of the resonance as proposed by  \citet{2016SerAJ.193...19M,2016ApJ...816L..31M}, where the strength of J7/3 MMR can be found in \citet{2006Icar..184...29G}. After some attempts, we finally set $T=0.02\,{\rm AU}/|\dot{a}_Y|$ empirically, which ensures that almost all of test particles leave the resonance in the end in our simulations. We note that the integration time $T$ for the adopted $|\dot{a}_Y|$ in this paper will be from 40\,Myr to 1\,Gyr. 

The software package \emph{Mercury6} \citep{1999MNRAS.304..793C}, with an additional Yarkovsky force subroutine, is adopted to simulate the orbits.
The \emph{hybrid} integrator is used, which is basically a symplectic integrator switching to Bulirsch-Stoer only during planetary close encounters. The Yarkovsky force is included in the subroutine \emph{mfo\_user} with the same method as \citet{2019A&A...622A..97Z}.

\section{Results} \label{sect:result}

Yarkovsky effect (mainly the diurnal effect) may increase the semi-major axis if the asteroid has a prograde rotation (the spinning obliquity $\gamma$ smaller than $90^\circ$), or drive the asteroid toward the Sun for the retrograde rotator ($\gamma > 90^\circ$). We simulate the outward migration and crossing of the J7/3 MMR for asteroids from two sets of initial conditions $Kor_{in}$ ($i_0=3^\circ$) and $Eos_{in}$ ($i_0=11^\circ$) both from $a_0=2.948$\,AU. And other two sets $Kor_{out}$ and $Eos_{out}$ both from $a_0=2.962$\,AU but with $i_0=3^\circ$ and $11^\circ$ respectively are for the inward migration and crossing of the resonance.

A test particle should migrate 0.02\,AU at the end of our numerical integration if it did not feel any resistance. But the J7/3 MMR is able to trap asteroids (though temporarily), thus some test particles (in fact very few) might still stay in the resonance until the end of integration. For those not in the resonance, some have been ejected from the neighbourhood via close encounters with planets after their orbits were excited by the resonance, and the rest of them cross the resonant region successfully and continue their migration.  

\begin{figure}[htbp]
   \centering
   \includegraphics[width=0.45\textwidth]{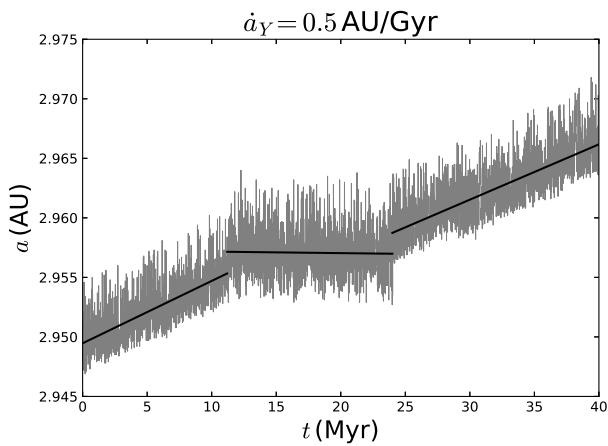}
      \caption{An example of resonance crossing. The black lines are piecewise linear fitting to guide eyes. }
         \label{fig:sample}
   \end{figure}

Generally we may tell whether an object is in the resonance by checking the resonant angle, but for high-order resonance like the J7/3 MMR in a complicated dynamical model, the critical angle seldom keeps oscillating all along. Instead, we adopt an indirect method similar to that in \citet{2016ApJ...816L..31M} to check if a test particle has crossed over the resonance safely or is still in the resonance. The key point of this method lies in the fact that the J7/3 MMR must modify the migration speed of test particles. We present in Fig.\,\ref{fig:sample} an example of the semi-major axis evolution of a test particle to show  our method of identifying the resonance crossing. 

As shown in Fig.\,\ref{fig:sample}, three stages of evolution can be seen clearly. When $t\lesssim 11$\,Myr before entering the resonance and $t\gtrsim 24$\,Myr after escaping from the resonance, averagely the semi-major axis drifting rate is just the given Yarkovsky drifting rate $\dot{a_Y}=0.5$\,AU/Gyr, i.e. the migration is determined solely by the Yarkovsky effect. While in between these two stages, the semi-major axis is nearly constant, implying that the J7/3 MMR takes control of the motion. 

Therefore, we monitor the semi-major axis drifting rate in our numerical simulations, and fit the time series of semi-major axis using the least squares method in a running window with width of 2\,Myr, which is 1/20 of the shortest integration time in our simulations. The test particles always start to migrate at the given speed $\dot{a_Y}$. Their migration will then be suppressed by the resonance after entering into the J7/3 MMR, in which the orbits might be excited and ejected. Except only a few ones staying in the resonance all along thereafter, most survivors from the J7/3 MMR will continue the migration at the same given speed $\dot{a_Y}$. As soon as the drifting rate was found to return back to the given $\dot{a_Y}$, the test particle is considered to have successfully crossed the J7/3 MMR. 

\subsection{Fraction of resonance crossing} \label{sect:cr}

\begin{figure}[htbp]
	\centering
	\includegraphics[width=0.45\textwidth]{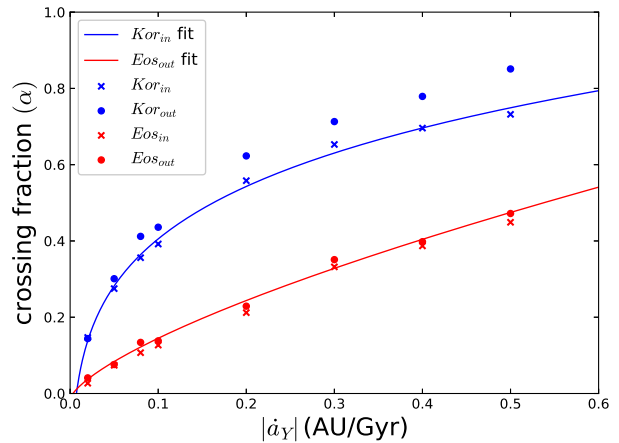}
	\caption{The crossing fraction against Yarkovsky drifting rate $|\dot{a}_Y|$. 
		Crosses and solid circles are for outward and inward migration respectively, while blue and red indicate initial inclination $i_0=3^\circ$ and $11^\circ$. Lines are fitting curves (see text).
	}
	\label{fig:cr_adot}
\end{figure}

In our simulations for four sets of initial conditions of test particles, we record the number of orbits that have made successful crossing by the end of numerical integration, and we summarize all the results in Fig.\,\ref{fig:cr_adot}. Eight values of $|\dot{a}_Y|$ from 0.02\,AU/Gyr to 0.5\,AU/Gyr are adopted, as mentioned previously. The positive $\dot{a_Y}$ for outward migration and negative $\dot{a_Y}$ for inward migration, and cases for two initial inclinations are plotted in the same figure. 

Apparently in Fig.\,\ref{fig:cr_adot} the crossing fraction increases with increasing $|\dot{a}_Y|$. Asteroids with large drifting rates cross the resonance easily, which is a foregone conclusion \citep{2006PhDT.......281B}.  Another interesting feature in Fig.\,\ref{fig:cr_adot} is the different crossing fractions for  opposite migrating directions. The transiting from the outer side to the inner side of J7/3 MMR (of $Eos_{out}$ and $Kor_{out}$) is always easier than the inverse transiting ($Eos_{in}$ and $Kor_{in}$). The difference in crossing fraction of opposite directions for Koronis family (low inclination) is much more significant than that for Eos family (high inclination). 
This difference is due to the asymmetry of dynamical structure of J7/3 MMR \citep{2003Icar..166..131T}, and the degree of asymmetry is higher at low inclination ($Kor_{in}$ and $Kor_{out}$) than that at high inclination ($Eos_{in}$ and $Eos_{out}$). Additionally, test particles at low inclination cross the resonance more easily than those at high inclination, because the dynamical structures of J7/3 MMR at different inclinations are different \citep{2003Icar..166..131T}. 

The relationship between the crossing fraction $\alpha$ and the Yarkovsky drifting rate $|\dot{a}_Y|$ can be numerically fitted using a function as 
\begin{equation} \label{eq:alpha}
\alpha=  C_1\left(\left|\dot{a}_Y\right|^{C_2}+C_3\right),
\end{equation} 
where $C_k$ $(k=1,2,3)$ are pending constants. For a better vision, we plot in Fig.\,\ref{fig:cr_adot} only the fitting curves for results of $Kor_{in}$ and $Eos_{out}$, which are two sets of test particles referring to the real asteroids of Koronis family and Eos family, respectively. Later in this paper, these functional relationships will be used to estimate the supplement flux of family members into the J7/3 MMR.

\subsection{Orbital Excitation} \label{sect:exc}

The orbits of asteroids may be excited by the combined effects of the resonance and the Yarkovsky effect. Currently, some asteroids have been found in the J7/3 MMR although it has been proven to be dynamically unstable. The orbital state of these asteroids is attributed to their sources and the excitation processes. 

\begin{figure}[htbp]
	\centering
	\includegraphics[width=0.45\textwidth]{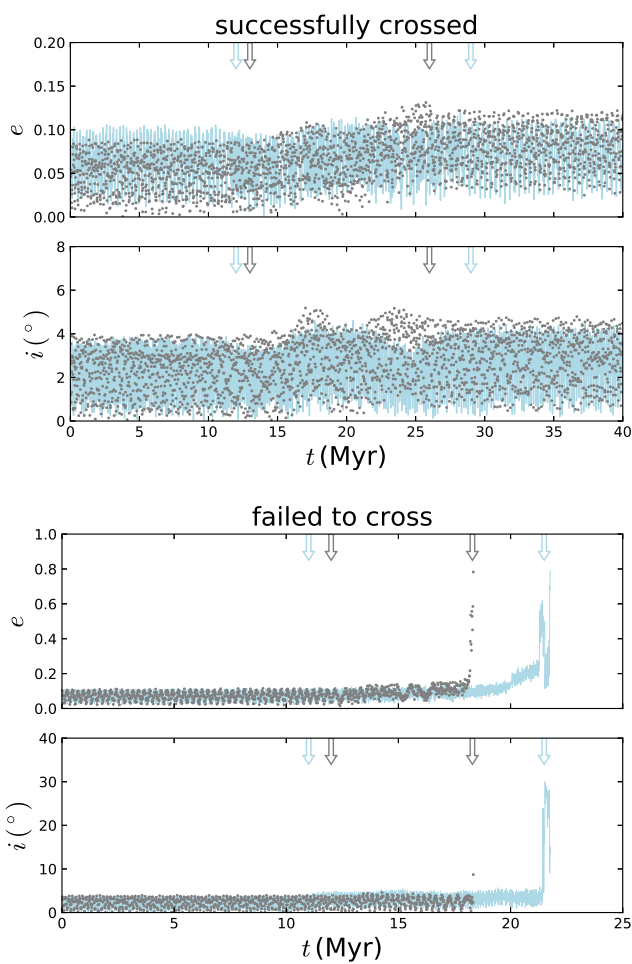}
	\caption{The evolutions of eccentricity and inclination for two asteroids that successfully crossed the resonance (upper two panels) and two asteroids that failed to cross the resonance (lower two panels). In each panel two objects are distinguished by colours. The arrows in corresponding colour denote the the moments of entering and leaving the resonance. All examples are randomly selected from the set of orbits $Kor_{in}$ and the Yarkovsky drifting rate is $\dot{a}_Y =0.5$\,AU/Gyr. 
	}
	\label{fig:ei_t}
\end{figure}

Checking the orbital evolutions of test particles in the simulations, we find that the objects that make successful crossing through the J7/3 MMR have very different behaviours from those that fail to cross. Four randomly selected but typical cases of the evolutions of eccentricity and inclination are presented in Fig.\,\ref{fig:ei_t}, two for objects that successfully crossed the resonance, and other two for objects that failed to cross the resonance. As shown, for the former case, both the eccentricity and the inclination are barely excited, but for the latter case the eccentricities are evidently pumped up, especially during the late stage of the resonance regime. Seemingly, the inclinations increase considerably but this excitation is mainly due to close planetary encounters after the objects have obtained the high eccentricity and left the resonance.

\begin{figure}[htbp]
	\centering
	\includegraphics[width=0.45\textwidth]{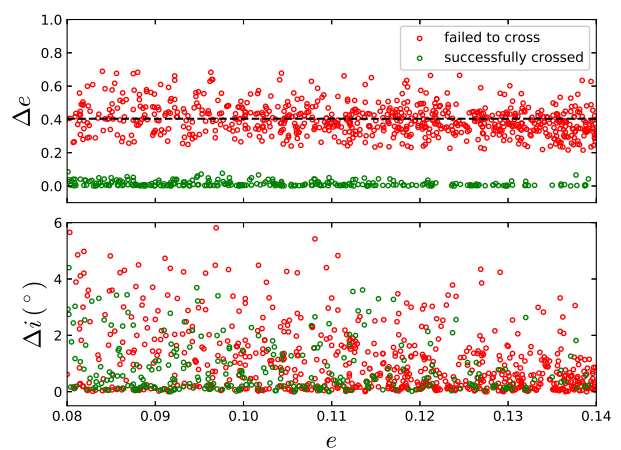}
	\caption{The excitations of eccentricity $\Delta e$ (top) and inclination $\Delta i$ (bottom) against initial eccentricity $e$ for initial set $Kor_{in}$. The Yarkovsky drifting rate $\dot{a}_Y=0.05$\,AU/Gyr. Green dots refer to test particles that successfully crossed through the J7/3 MMR, while red dots for those that failed to. The dashed line represents the average value of red dots.
	}
	\label{fig:dedi_e}
\end{figure}

To find out the statistical results about the dynamical excitation, we reckon the variations of the eccentricity and inclination of test particles in all our simulations.  For asteroids successfully crossing through the resonance, a statistical estimation of the variations of eccentricity and inclination can be obtained by calculating the difference between the maximum and the minimum of the corresponding mean orbital element ($\Delta e$ and $\Delta i$) during the evolution. The mean orbital elements are derived in a running window with width of 0.2\,Myr. For those asteroids that failed to cross the resonance, being either trapped in or ejected out of the resonance, in order to reveal the dynamical influences solely from the combined  Yarkovsky effect and the J7/3 MMR, we should eliminate the part of excitation that attributes to close encounters with planets. To do so, we expel the orbital evolution since a short period before the first-time-ever close encounter with a planet (defined as when the asteroid is within 3 Hill radii from the corresponding planet), which is reported by the integrator package \emph{Mercury6}. Then the $\Delta e$ and $\Delta i$ are calculated just as in the former case of successful crossing. We summarize the results as follows and present as an example the case for the initial set $Kor_{in}$ in  Fig.\,\ref{fig:dedi_e}.

As shown in Fig.\,\ref{fig:dedi_e}, the variations of eccentricity obviously split into two distinct groups, one for test particles that successfully crossed the J7/3 MMR and the other one for those failed. In the former case, the excitation of eccentricity is less than 0.1, while the eccentricities in the latter case increase by about 0.4 in average, with the  maximal $\Delta e$ being larger than 0.7, which is consistent with the result in \citet{1995Icar..114...33M}. In these extreme cases, asteroids with eccentricity up to 0.7 or higher will have a chance to become NEOs directly, without any close encounters with other planets. 

In addition, it seems that there is no evident relationship between the excitation and initial value of eccentricity. And surprisingly, no test particle appears in the intermediate zone between the above mentioned groups, implying that the Yarkovsky drift and the resonance contribute to the orbital excitation in a complicated way. The excitation mainly attributes to the resonance. In fact, without the Yarkovsky drift, the eccentricity of asteroids in the J7/3 MMR undergoes slow chaotic diffusion for millions of years and this process is then followed by a fast increase and large oscillations of eccentricity until the asteroid is scattered away \citep{2003Icar..166..131T}. After the Yarkovsky drift is introduced, such dynamical evolution route may be well kept for some objects (see Fig.~\ref{fig:ei_t}, bottom two panels), thus their eccentricities are significantly pumped up and they fail to cross through the resonance.  
	
There is also a certain probability that the forced migration may interrupt the above mentioned process, preventing the asteroids from entering the quick excitation phase, and finally convey the asteroids through the J7/3 MMR. In this case, the eccentricity excitation arises from the slow chaotic diffusion as well as the Yarkovsky migration.

\citet{2008MNRAS.390..715B} found that the Yarkovsky effect might drive the eccentricity of a prograde spinning asteroid locked in a first-order MMR  to increase, or decrease if the asteroid was a retrograde rotator. Similar phenomenon has been observed, e.g. in the 3/2 MMR with Jupiter \citep{2017Icar..288..240M} and in the 1/1 MMR with the Earth  \citep{2019A&A...622A..97Z}.  But the J7/3 MMR studied in this paper, as a 4th-order resonance and under the influences of complicated perturbations from other planets in the dynamical model, is relatively weak in term of dynamical effect. Therefore, the mechanism of transferring the Yarkovsky effect on semimajor axis to the eccentricity \citep{2008MNRAS.390..715B} works less effectively here, and the eccentricity excitation is very limited. Of course, the interaction between the Yarkovsky effect and the weak MMRs deserves a thorough investigation in future. 

As for inclination, the separation between two groups disappears. Test particles trapped in the resonance have low excitation in inclination and are mixed up with those crossed through the resonance, as shown in the lower panel of Fig.\,\ref{fig:dedi_e}. This happens because neither the J7/3 MMR nor the Yarkovsky effect is able to pump up the inclination. This result is consistent with the conclusion in \citet{2017A&A...598A..52G} that the distribution of inclinations for J7/3 MMR is virtually constant.

The orbital excitations in simulations for other sets of test particles are similar to the ones shown in Fig.\,\ref{fig:dedi_e}. The Yarkovsky drift rate and the drift direction have only very limited influence on the orbital excitation. 
If a test particle safely crosses through the J7/3 MMR, its eccentricity will increase about 0.03 in average, and its inclination will increase about $0.5^\circ$ in average. For those test particles that failed to cross the J7/3 MMR, the excitation in eccentricity is about 0.4 averagely, while for inclination it increases by about $1^\circ$ in average. 

\subsection{Escaping from resonance} 

Starting from the initial positions, all test particles in our simulations will migrate toward then meet with the J7/3 MMR. After running into the J7/3 MMR, a test particle spends a period of time before being ejected or making a successful crossing. We follow the evolution of each orbit and divide all particles into two groups according to their final destinations, one group for those particles that are ejected by the resonance and the other one for particles that cross over the resonance successfully. As mentioned in Section\,\ref{sect:exc}, those ejected particles must have been excited by the resonance. Thus, their destiny is mainly dominated by the resonance, and we call this group of particles \emph{Resonant Group}. 
Furthermore, those particles that finally cross the J7/3 MMR successfully constitute the \emph{Transit Group}.
Particles in both groups will leave the J7/3 MMR, and below we discuss their escaping from the resonance separately. 

\begin{figure}[htbp]
	\centering
	\includegraphics[width=0.45\textwidth]{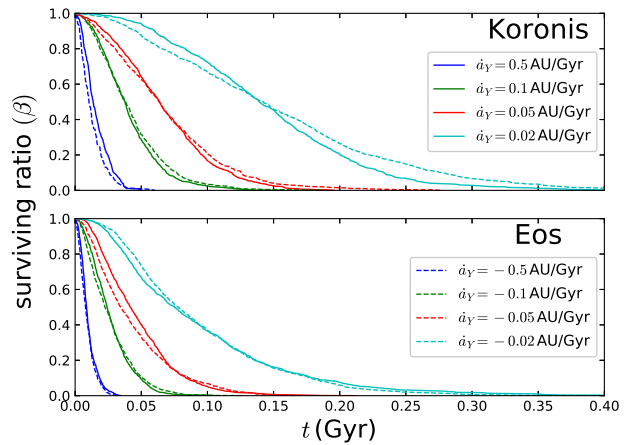}
	\caption{The surviving ratio against time for test particles in \emph{Resonant Groups}. Different Yarkovsky drifting rates $|\dot{a}_Y|$ are adopted for four sets of initial conditions. The upper panel labelled {\it Koronis} is for two sets with Koronis-like inclination $i_0=3^\circ$ while the lower panel is for Eos-like inclination sets with $i_0=11^\circ$. The solid and dashed lines indicate the outward and inward migration by Yarkovsky effect, respectively.
	}
	\label{fig:sr_trap}
\end{figure}

As particles in both groups escape from the J7/3 MMR, the number of particles that remain in the resonance region decreases with time. For Resonant Group, without Yarkovsky effect the decreasing rate reflects purely the dynamical stability of this resonance \citep{2003Icar..166..131T}. After considering the Yarkovsky effect, additional decreasing arises mainly because the Yarkovsky effect influences the stability, as well as because a few particles may be lost when they are forced to cross the boundary of this resonance, where a broad chaotic zone associated with the resonance separatix and the accumulated secondary resonances \citep[e.g.][]{henrard1990, 1990Icar...85..444M, ferraz1996} is located. 
We summarize the decreasing of surviving ratio (normalized number of particles) in the Resonant Groups with respect to time in Fig.\,\ref{fig:sr_trap}.

Since the initial positions of test particles are all out of the resonance, they spend some time in approaching the resonance. We disregard this transitional period and set the moment of the first escape of test particle as the starting point of timing. It should be noted that this time-zero is not exactly the moment when the test particle reaches the boundary of J7/3 MMR, which is almost impossible to locate accurately, but the first escape happens very soon after the particles reach the resonance. Thus, such selection of time zero is good enough for our purpose of studying the decaying of surviving ratio with time, which happens in the order of tens of millions years. 

As illustrated in Fig.\,\ref{fig:sr_trap}, the surviving ratio for Eos-like particles (lower panel) drops more quickly than Koronis-like particles (upper panel). This is because the J7/3 MMR of $\sim$10$^\circ$ is less stable than that of $\sim$3$^\circ$ \citep{2003Icar..166..131T}. The Yarkovsky effect enhances the instability, shortens the dynamical lifetime of particles. The larger the Yarkovsky drifting rate, the shorter the typical lifetime. In addition, the migration direction, either inward or outward, makes no evident difference in the decreasing rate, not like in the case of Earth Trojans under the influence of Yarkovsky effect, where the inward migration increases the libration amplitude of the resonant angle of the 1/1 MMR and {\it vice versa} \citep{2019A&A...622A..97Z}. 

\begin{figure}[htbp]
	\centering
	\includegraphics[width=0.45\textwidth]{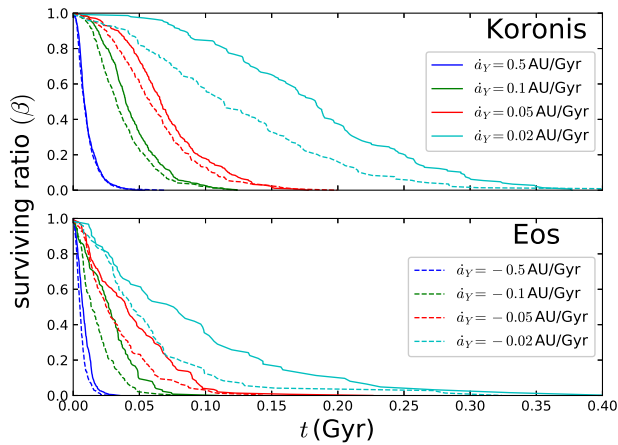}
	\caption{Same as Fig.\,\ref{fig:sr_trap} but for test particles in \emph{Transit Groups}. 
	}
	\label{fig:sr_cross}
\end{figure}

The Transit Group is consisted of particles that make successful crossing of the resonance. Unlike the particles in the Resonant Group that escape from the resonance region and get removed ``violently'' in excited orbits, particles in the Transit Group leave the resonance by crossing the boundary ``peacefully'' with relatively small eccentricities. For Transit Group, the decreasing of  the numbers of particles that still remain in the J7/3 MMR with respect to time is presented in Fig.\,\ref{fig:sr_cross}. 
It is worthy to note that the J7/3 MMR does impede the transition of these particles, otherwise for a given Yarkovsky drifting rate they should cross the resonance all at the same moment because they all start from the same initial semi-major axes as we set. Also note that the total number of particles in the Transit Group for slow Yarkovsky migration (small $|\dot{a_Y}|$) is small, because the crossing fraction is small for slow migration (see Fig.\,\ref{fig:cr_adot}). 

The contours of the curves in Fig.\,\ref{fig:sr_cross} are similar to the ones in Fig.\,\ref{fig:sr_trap}. And similarly, the larger Yarkovsky drifting rates lead to quicker decreasing of number of particles. But different from Fig.\,\ref{fig:sr_trap}, in Fig.\,\ref{fig:sr_cross} the solid curves for outward migration are always above the dashed curves for inward migration, implying that the prograde spinning asteroids (thus $\dot{a_Y}>0$, outward migration) will leave the J7/3 MMR in a slower pace than the retrograde rotators. 

In a dynamical system the surviving ratio of orbits generally decays with time following either an exponential law in the fully chaotic region of phase space, or an algebraic law in the mixed region of chaotic and regular motion \citep[see e.g.][p.147]{sun2015}, but we find the curves in Fig.\,\ref{fig:sr_trap} and Fig.\,\ref{fig:sr_cross} can be well fitted by Gaussian function as follows,
\begin{equation} \label{eq:beta}
\beta=\exp\left[-\left(\frac{t-t_1}{t_2}\right)^2\right].
\end{equation} 
Here we denote the surviving ratio by $\beta$, and $t_1, t_2$ are pending parameters. Obviously, $t_1$ is the time after which the test particles begin to escape from the resonance (either being ejected or just cross the resonance boundary), and $t_2$ indicating the width of the fitting Gaussian curve is the e-folding time. 
The parameters $t_{1,2}$ depend on the Yarkovsky drifting rate $|\dot{a}_Y|$, and the dependence can be numerically fitted using the same function as in Equation\,\eqref{eq:alpha}. Finally, the surviving ratio as a function of time $t$ and the drifting rate $|\dot{a}_Y|$ reads
\begin{equation} \label{eq:beta2}
\beta(t, |\dot{a}_Y|)=\exp\left\{-\left[\frac{t-\left(p_{11}|\dot{a}_Y|^{p_{12}}+p_{13}\right)}{p_{21}|\dot{a}_Y|^{p_{22}}+p_{23}}\right]^2\right\}
\end{equation}
where $p_{kj}$ $(k=1,2; j=1,2,3)$ are numerically fitting coefficients. For each set of initial conditions of test particles ($Kor_{in}, Kor_{out}, Eos_{in}, Eos_{out}$), these coefficients are calculated and the functional relationship $\beta(t, |\dot{a}_Y|)$ are obtained. We should note that this is a purely empirical function and the physical or dynamical meaning of these coefficients is ambiguous. However, the function can be applied to make useful estimation of the supplement flux of asteroids to the J7/3 MMR, as we will show in next section. 
   
\section{Transportation of asteroids by J7/3 MMR } \label{sect:app}
The results presented above should be verified by observational data. After that, we will apply them to investigate the replenishment of asteroids in the J7/3 MMR by members of Koronis family and Eos family.

\subsection{Asteroid family members around J7/3 MMR} \label{sect:res-fmly}

The Koronis family and Eos family originally reside in the inner and outer side of the J7/3 MMR respectively. Assisted by the Yarkovsky effect, members of these two families may approach and cross the resonance from opposite directions. As shown by our simulations in previous section, those asteroids in {\it Transit Group} may keep their original eccentricities and inclinations after crossing the J7/3 MMR. Thus, they might still be recognised as the members of the family with the hierarchical clustering method introduced in \citet{1990AJ....100.2030Z}. 

\begin{figure}[htbp]
   \centering
   \includegraphics[width=0.45\textwidth]{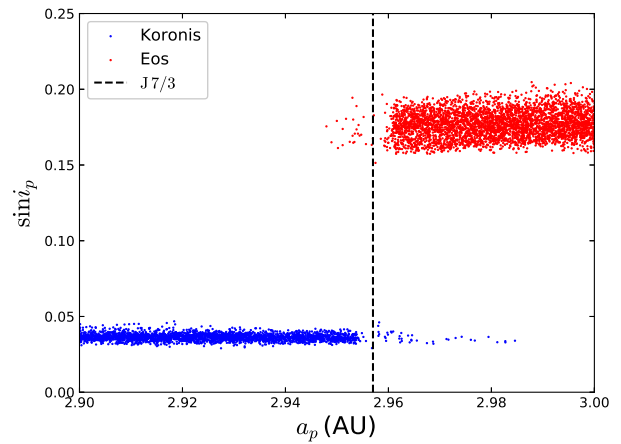}
      \caption{The distribution of members of Koronis family (blue dots) and Eos family (red dots) around the J7/3 MMR. The dashed line shows the location of the center of J7/3 MMR. 
                       }
         \label{fig:J7/3}
   \end{figure}

We download the data of members of these two families from \emph{AstDyS}\footnote{\url{http://newton.spacedys.com//astdys/}} \citep{2014Icar..239...46M} and show the proper semi-major axes and inclinations in Fig.\,\ref{fig:J7/3}. Apparently, the J7/3 MMR serves well as a boundary to both families, but still a few members that have transited through the resonance can be found. The numbers of these transited members can be used to set constraints on the possible evolution history of these families.  

\begin{figure}[htbp]
	\centering
	\includegraphics[width=0.50\textwidth]{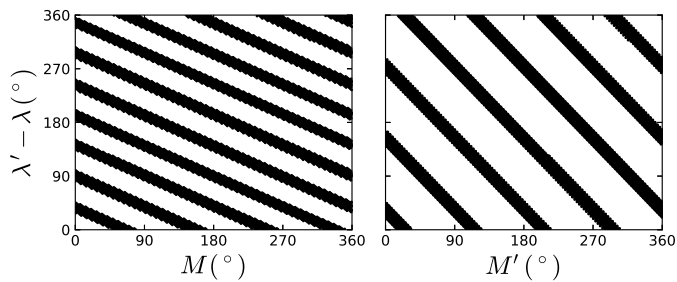}
	\caption{An example of the FAIR method. The asteroid is  (348975) 2006 UG111, an Eos family member. $\lambda$ and $M$ are mean longitude and mean anomaly of the asteroid, while the prime denotes the ones of Jupiter.   
	}
	\label{fig:fair}
\end{figure}

In fact, asteroids found near the exact location of the J7/3 MMR in Fig.\,\ref{fig:J7/3} might be  trapped in the resonance currently, which should be identified first. The FAst Identification of mean motion Resonance (FAIR) method by \citet{2018MNRAS.477.3383F} is adopted here to identify these resonant asteroids. We numerically simulate the orbital evolutions of asteroids near the J7/3 MMR for $10^4$ years, and plot the scatter diagrams of some critical angles, which help us identify the resonance. An example is shown in Fig.\,\ref{fig:fair}. 

In Fig.\,\ref{fig:fair}, we see 7 centres along the abscissa and 4 centres along the ordinate in the $\lambda^\prime-\lambda$ versus $M$ plot, while in the $\lambda^\prime-\lambda$ versus $M^\prime$ plot the numbers of centres are 3 and 4, indicating that this asteroid is in the 7/3 MMR with Jupiter \citep[for more details please refer to ][]{2018MNRAS.477.3383F}. For an asteroid not in the resonance, these scatter plots will be just a mess. 

From \emph{AstDyS} website, we download the data of 829 asteroids around the J7/3 MMR with semi-major axes being between 2.952\,AU and 2.962\,AU. Via the FAIR method, we recognise 440 asteroids currently trapped in the resonance, among which 66 and 26 have been identified as the members of Koronis family and Eos family, respectively. 

In addition to these 92 ($=66+26$) family members trapped in the J7/3 MMR, we find 24 Koronis asteroids on the right hand side of this resonance and 6 Eos asteroids in the left hand side (see Fig.~\ref{fig:J7/3}). They are the family members that have crossed successfully the J7/3 MMR. However, some calibration should be made before these numbers can be used to give meaningful estimations.

\subsection{Crossing fraction of family members} \label{sect:comp}
\citet{2013A&A...559A.134H} suggest that the centres of Koronis and Eos family locate at $a_K\approx2.870\,$AU and $a_E\approx3.015\,$AU, respectively. For the ages of these two families, we notice that some different values have been proposed in literature. For example, we may find the age of Koronis family $T_K=2.5\pm1.0\,$Gyr \citep{2001Sci...294.1693B} and $T_K=2.5\sim4.0\,$Gyr \citep{2013A&A...559A.134H}, while for Eos family $T_E=2.0\pm0.5\,$Gyr \citep{2005Icar..173..132N}, $1.3^{+0.15}_{-0.2}\,$Gyr  \citep{2006Icar..182...92V} and $1.3\pm0.5\,$Gyr \citep{2015aste.book..297N}. Based on these references, we adopt roughly the ``averaged'' value of family age $T_K=2.5\,$Gyr and $T_E=1.5\,$Gyr. 

It is not easy to determine the exact location of the boundary of J7/3 MMR. We note that to the left of J7/3 MMR, the largest semi-major axis among the non-resonant members from both Koronis and Eos families is $a_1=2.951$\,AU, and this value is approximately set as the left border of J7/3 MMR. In a similar way, we define the smallest semi-major axis of the non-resonant family members to the right of J7/3 MMR as the right border $a_2=2.962$\,AU. Starting from the birth place (the centre of the family), a member of Koronis family has to travel $|a_1-a_K|$ to reach the J7/3 MMR. Therefore, those Koronis family members who are reaching the J7/3 MMR nowadays have approximately the migration speed of $\dot{a}_{YK}=(a_1-a_K)/T_K$. The same calculation gives the migration speed for Eos family members approaching now the resonance as $\dot{a}_{YE}=(a_2-a_E)/T_E$. Adopting the resonance borders $a_1,a_2$, family centres $a_K, a_E$ and family ages $T_K, T_E$, we estimate the Yarkovsky drifting rate of Koronis and Eos asteroids that are entering the J7/3 MMR as $\dot{a}_{YK}=0.0353$\,AU/Gyr and $\dot{a}_{YE}=-0.0324$\,AU/Gyr, respectively. Approximately, these drifting rates can be regarded as the typical migration speeds in recent period of time for the two families' members that are now located around the J7/3 MMR. 

The leftmost Eos member in Fig.\,\ref{fig:J7/3} has the minimum of semi-major axis  $a_{\min}=2.948\,$AU. With the migration speed mentioned above, it has spent $\Delta T = (a_{\min}-a_1)/\dot{a}_{YE} = 0.085$\,Gyr in travelling after it crossed the J7/3 MMR. During the same period, the Koronis asteroid escaping from the J7/3 MMR should have travelled to a position $a_{\max}= a_2+\dot{a}_{YK}\Delta T = 2.965$\,AU. Consequently, Eos asteroids in $(a_{\min}, a_1)$ and Koronis asteroids in $(a_2,a_{\max})$ crossed the resonance during the same period of time $\Delta T$, and their numbers reflect the flux of family members crossing the J7/3 MMR. From the observational data, we find 9 Koronis asteroids and 6 Eos asteroids in the aforementioned semi-major axis intervals, and the Koronis-Eos ratio is $1.50$ $(=9/6)$. 

Adopting the Yarkovsky drifting rates $0.0353$\,AU/Gyr and $-0.0324$\,AU/Gyr for Koronis and Eos asteroids, we calculate the crossing fraction $\alpha$ using Equation~\eqref{eq:alpha}. The results are $\alpha_K=0.211$ and $\alpha_E=0.0626$ for two populations. The ratio $\alpha_K/\alpha_E=3.37$ is much higher than 1.50 because we have assumed that equal number of asteroids approach the J7/3 MMR per time from both families. In fact, there are about $n_K=87$ Koronis asteroids and $n_E=196$ Eos asteroids per 0.002\,AU around the J7/3 MMR. Taking into account of the number density of asteroids and the Yarkovsky drifting rates together, we obtain the flux ratio between these two populations $(n_K\dot{a}_{YK})/(n_E\dot{a}_{YE})=0.407$, and the final Koronis-Eos crossing ratio is $3.37\times 0.407 = 1.37$. This number (1.37) agrees roughly with the ratio from observational data (1.50), although the statistical significance of the latter is limited by small sample numbers. This agreement verifies our calculations so far. 

It is worthy to note that the estimation given above may be affected by several uncertainties, including the ages of two families, the adopted Yarkovsky drifting rates, and the membership recognition of the family members particularly those strongly influenced by the resonance. 

Because the resonance may significantly change the orbital characteristics of an asteroid, it requires special caution to identify the family members on the opposite side of the resonance with respect to the family centre. In this sense, the global method applied to the whole main belt population \citep{2014Icar..239...46M}, by which we obtain the families' members in this paper, may not be the best choice. We examine the data of the Koronis and Eos families on the \emph{AFP}  \citep[Asteroid Families Portal, ][]{2017MNRAS.470..576R} website\footnote{\url{http://asteroids.matf.bg.ac.rs/fam/index.php}}, in which some interlopers are removed, and we find in this data that there is no Eos member on the left hand side of J7/3 MMR and only 5 Koronis members reside on the opposite side. Obviously, more observations with less bias (favouring in big asteroids thus slow Yarkovsky drift) and more careful family membership recognition are needed in future.

\subsection{Replenishment of asteroids in J7/3 MMR}

We find 440 asteroids in the J7/3 MMR, among which 92 are family members. Since the J7/3 MMR is dynamically unstable, these resonant asteroids must be replenished constantly from some sources, two of which are Koronis family and Eos family \citep{2003Icar..166..131T}. 

Using the fitting functions in Section~\ref{sect:result}, we are able to calculate the amount of replenishment from these two asteroid families by the formula
\begin{eqnarray} \label{eq:numint}
\begin{aligned}
N_{7/3} = \int_{-\infty}^{t} F(\tau) & \bigg[ \Big( 1-\alpha(\tau)\Big)\,\beta_R\left(t-\tau,|\dot{a}_Y|\right) \\
& + \alpha(\tau)\,\beta_T\left(t-\tau,|\dot{a}_Y|\right)  \bigg] {\rm d} \tau.
\end{aligned}
\end{eqnarray}
where $N_{7/3}$ is the number of asteroids in the resonance, $F$ is the flux of asteroids approaching to the starting location where our simulations in Section~\ref{sect:result} were started, $\tau$ is the time needed for an asteroid to travel from its orbit to the starting location, $\alpha$ is the crossing fraction given in Equation~\eqref{eq:alpha}, $\beta_R$ and $\beta_T$ are the surviving ratio of asteroids in Resonant Group and Transit Group given both by the function in Equation~\eqref{eq:beta2} but with different fitting coefficients according to curves in Fig.\,\ref{fig:sr_trap} and Fig.\,\ref{fig:sr_cross}.  

\begin{figure}[htbp]
	\centering
	\includegraphics[width=0.5\textwidth]{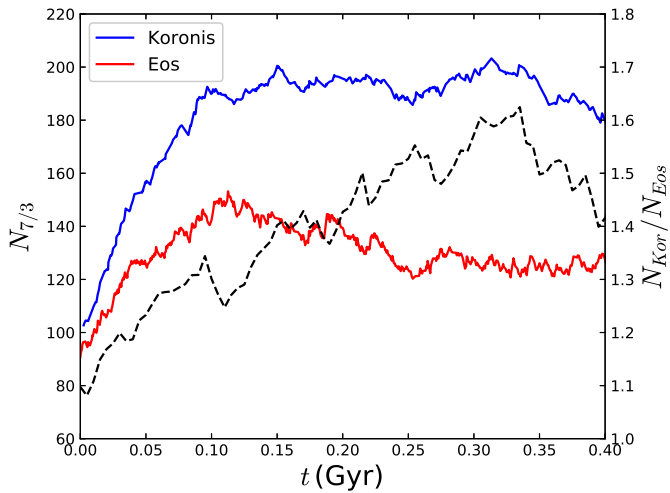}
	\caption{The numbers of asteroids in the J7/3 MMR attributed by Koronis family (blue) and Eos family (red). The dashed line is the ratio between these two populations, specified on the right ordinate axis. The initial time $t=0$ corresponds to MJD\,58200.
	}
	\label{fig:res}
\end{figure}

In practice, the integral in Equation\,\eqref{eq:numint} can be replaced by a summation as follow. 
\begin{equation} \label{eq:numsum}
N_{7/3} = \sum_{-\infty <\tau\le t}\,\bigg [\Big (1-\alpha(\tau)\Big )\,\beta_R\left(t-\tau,|\dot{a}_Y|\right) 
+ \alpha(\tau)\,\beta_T\left(t-\tau,|\dot{a}_Y|\right)\bigg ].
\end{equation}
In this summation, $F\Delta\tau=1$ by definition, where $\Delta\tau$ is the difference of arrival time between two asteroids successively arriving at the starting location, and that's why $F$ and $\Delta\tau$ do not appear explicitly. In calculating this summation, the time delay $\tau$ of an asteroid can be calculated using the distance from its current orbit to the starting location (in semi-major axis) and the migrating speed ($|\dot{a}_Y|$), which in turn is simply approximated by the quotient of its distance to the family centre divided by the family age. Adopting the data of family members downloaded from \emph{AstDyS}, we calculate for each asteroid the values of $|\dot{a}_Y|$, $\tau$ and the probabilities $\alpha, \beta_R, \beta_T$. The results obtained from Equation~\eqref{eq:numsum} are shown in Fig.\,\ref{fig:res}. At the end of our calculation ($t=0.4$\,Gyr from MJD\,58200), 1321 Eos members and 626 Koronis members from the \emph{AstDyS} data have been fed into the summation.

It should be noted that the lower limit of integration in Equation~\eqref{eq:numint} is set as $-\infty$ to emphasize that theoretically the calculation should start from the ``real'' initial time in the past. However, the original statuses of the two asteroid families are not known, thus we have to perform the practical calculation of Equation~\eqref{eq:numsum} based on the current data, including those family members now trapped in the J7/3 MMR. 

As shown in Fig.\,\ref{fig:res}, after a transient period due to the lack of knowledge of the past, the replenishment of asteroids from Koronis and Eos family may maintain an almost steady number of about 320 asteroids in the J7/3 MMR, among which $\sim$190 from Koronis and $\sim$130 from Eos family. Since the current population in the J7/3 MMR is the legacy of the asteroids flux in the past, which is not known, we will not compare directly this total number (320) with the observational data. In fact, two additional reasons as well stop us from making direct comparison. One is the J7/3 MMR may get supplemented by capturing background asteroids (but not necessarily from the two families), and the other reason is that even if the replenishments are the families' members their recognition as family members may be difficult due to the perturbation arising from the resonance. 

However, among the total 440 resonant asteroids found in the J7/3 MMR, the number ratio between the Koronis asteroids (66) and Eos asteroids (26) is higher than the value $N_{Kor}/N_{Eos}$ in our calculation (Fig.\,\ref{fig:res}), which is always smaller than 1.6. One possible explanation is that the J7/3 MMR takes a shorter time to modify the orbital characteristics of Eos asteroids than that of Koronis asteroids, thus the Eos asteroids are easier to lose their original identities as the source family members. That is to say, in the J7/3 MMR there are in fact more asteroids originated from Eos family than we know, we just cannot recognize them. 

\subsection{Transportation to near Earth region}
\begin{figure}[htbp]
   \centering
   \includegraphics[width=0.45\textwidth]{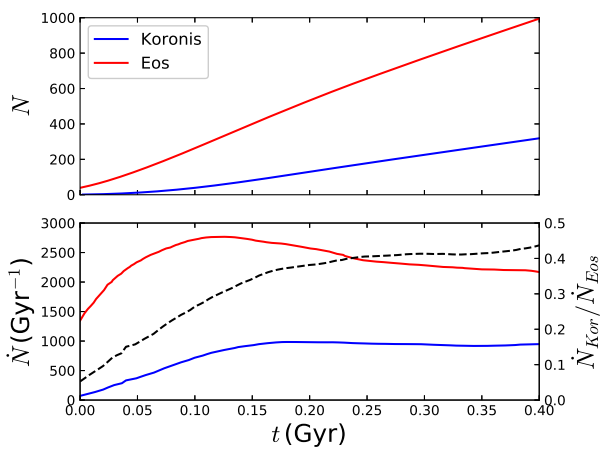}
      \caption{Accumulated number (top) and flux (bottom) of asteroids transported out via J7/3 MMR from Koronis (blue) and Eos (red) families. The dashed line is the ratio of transported asteroids from Koronis family to those from Eos family ($\dot{N}_{Kor}/\dot{N}_{Eos}$).
                       }
         \label{fig:trans}
   \end{figure}

The J7/3 MMR is continuously replenished by asteroids from both Koronis and Eos families via the Yarkovsky effect. And subsequently, some of them, i.e. those members that cannot make successful crossing of the resonance, will be scattered out from the J7/3 MMR and transported to other regions in the solar system, e.g. to become NEOs. Following the same strategy adopted above, we derive the formula describing the accumulated number of asteroids transported out via the J7/3 MMR as below.
 \begin{equation}
N = \sum_{-\infty<\tau\le t}\,\bigg[ \Big (1-\alpha(\tau)\Big )\,\Big (1-\beta_T\left(t-\tau,|\dot{a}_Y|\right)\Big ) \bigg],
\end{equation}
where $\alpha$ and $\beta_T$ are the same as in Equation~\eqref{eq:numsum}. And we present in Fig.\,\ref{fig:trans} the calculated flux of asteroids of both families transported out by the J7/3 MMR. 

Apparently, there are 2$\sim$3 times more Eos family members scattered out from the J7/3 MMR than that of Koronis family, despite the fact that more Koronis asteroids might be found in the resonance as shown in Fig.\,\ref{fig:res}. The flux from Koronis family to the J7/3 MMR is relatively smaller while Koronis asteroids have a higher crossing probability (see Fig.\,\ref{fig:cr_adot}), as a result, the ratio of Koronis asteroids to Eos asteroids reaches $\sim$0.4 after the transient period of time, as shown in Fig.\,\ref{fig:trans}. 
\citet{2017A&A...598A..52G} obtained a similar result that the ratio between asteroids with inclinations of $2^\circ$ and $10^\circ$ for J7/3 MMR is about 0.3. Since the objects in their study are the whole main belt asteroids rather than Koronis and Eos families only, the difference between their result and ours is acceptable in this degree.

It is important to bear in mind that due to the Yarkovsky effect asteroids pushed into the J7/3 MMR from Koronis family must be prograde spinning ones while those from Eos family are retrograde rotators. If we believe that asteroids from both families will evolve following the same rule after they escaped the J7/3 MMR, they will have the same probability of visiting any area in the solar system, including the near Earth region. Therefore, as a source of NEOs, our calculations show that the J7/3 MMR sends $2\sim 3$ times more retrograde spinning Eos family members to the NEO orbits than the prograde spinning Koronis family members, just as the accumulated numbers of the transported asteroids show in Fig.\,\ref{fig:trans}. If the inclinations of asteroids are conserved in the following evolution after escaping from the J7/3 MMR (the inclination is more unalterable than other orbital elements such as  semi-major axis and eccentricity), even more accurate comparisons with the NEO population can be made in future. 

\section{Conclusions} \label{sect:conclusion}
The mean motion resonances (MMR) are thought to be responsible for the formation of Kirkwood gaps in the main asteroid belt. Asteroids in these MMRs accumulate eccentricities until they are removed by close encounters with planets. Meanwhile, due to the Yarkovsky effect, asteroids are forced to migrate into and replenish these MMRs. In this paper, we investigate such transportation of asteroids originated from the Koronis family and the Eos family into the 7/3 MMR with Jupiter (J7/3 MMR). 

Our calculations show that from both families asteroids may either successfully cross through or be scattered away by the J7/3 MMR. The fraction of successful crossing is found to depend on the drifting rate and direction of the Yarkovsky migration, and also on the orbital inclination. Relatively, the quick and inward migration of asteroids with low inclination is in favour of the high probability of crossing the J7/3 MMR. 

Simultaneously driven by the Yarkovsky effect, the orbits of asteroids in the J7/3 MMR are excited by the resonance, and the excitation is found to be barely dependent on the strength and direction of the Yarkovsky effect, neither on the initial eccentricity and inclination of orbits.
      
From the J7/3 MMR, asteroids are removed continuously either because of the resonance that pumps up the eccentricities or because of the Yarkovsky effect that drives the asteroids to cross over the resonance. For both mechanisms, the surviving ratio of asteroids in the J7/3 MMR drops down with time, and the rate is determined mainly by the strength of Yarkovsky effect. The larger the Yarkovsky drifting rate is, the more quickly the surviving ratio drops.

The Koronis family resides on the left side of the J7/3 MMR, while the Eos family is on the right side. However, some family members from both families are found on the opposite side of the J7/3 MMR. We estimate the Koronis-to-Eos ratio of family members that successfully cross the resonance, and our calculations suggest a ratio of 1.37, which is in agreement with the observational data.

Although the J7/3 MMR is not a dynamically stable region for asteroids, we find from the observational data $\sim$440 asteroids that are on the 7/3 resonant orbits. The replenishment of asteroids into the J7/3 MMR is investigated. Our estimation based on the current knowledge of the families members shows that two asteroid families may provide a steady flux of asteroids into the resonance to maintain a population of $\sim$190 members from Koronis family and $\sim$130 members from Eos family in the coming 0.4\,Gyr. Of course, the background asteroids will also contribute to the resonant population. 

Aided by the Yarkovsky effect, the J7/3 MMR acts as a transportation route for asteroids from its vicinity to other regions in the solar system by exciting the orbits of asteroids that are injected into this resonance. Our calculations show that the transported number of Eos asteroids is 2$\sim$3 times that of Koronis asteroids. A fraction of these transported asteroids will arrive in the near Earth region, thus we may find among near Earth objects the footprints of this transportation process, such as the distribution of their orbital inclinations and their spinning axes' directions. 

Last but not least, we have adopted the simplified Yarkovsky effect model, in which only the semi-major axis of asteroids is modified monotonically and continuously. Although the collisions among the asteroids and the YORP effects must introduce randomness and uncertainties to the migration of individual asteroids, averagely and statistically the calculations in this paper are still valid.

\begin{acknowledgements}
Our great appreciations go to Dr. Bojan Novakovic for his comments and suggestions that helped us improve the manuscript. We also thank Dr. Kleomenis Tsiganis for his constructive comments during revising the paper. This work has been supported by the National Key R\&D Program of China (2019YFA0706601) and National Natural Science Foundation of China (NSFC, Grants No.11473016 \& No.11933001).      
\end{acknowledgements}

\end{document}